\documentclass[10pt,letterpaper,twocolumn]{article} 

\usepackage{ol2}
\usepackage[draft]{hyperref}
\usepackage{amsmath}
\usepackage{color}

\begin{document}

\twocolumn[ 
\hyphenation{pro-ba-bi-li-ty}
\title{Direct Schmidt number measurement of high-gain parametric down conversion}


\author{I.~V.~Dyakonov$^{1}$, P.~R.~Sharapova$^{1,2}$ T.~Sh.~Iskhakov$^{3,*}$, G.~Leuchs$^{3,4}$}

\address{
$^1$Physics Department, Moscow State University, Leninskiye Gory 1-2, Moscow 119991, Russia
\\
$^2$Skobeltsyn Institute of Nuclear Physics, Lomonosov Moscow State University, Moscow 119234, Russia\\
$^3$Max-Planck Institute for the Science of Light, Guenther-Scharowsky-Str. 1 / Bau 24, Erlangen  D-91058, Germany \\
$^4$University of Erlangen-N\"urnberg, Staudtstrasse 7/B2, 91058 Erlangen, Germany
\\

$^*$Corresponding author: Timur.Iskhakov@mpl.mpg.de
}

\begin{abstract} In this work we estimate the transverse Schmidt number for the bipartite bright squeezed vacuum state by means of second-order intensity correlation function measurement. Assuming that the number of modes is equal in both beams we determine the Schmidt number considering only one of the subsystems. The obtained results demonstrate that this approach is equally efficient over the whole propagation of the state from the near field to the far field regions of its emitter.

\end{abstract}

\ocis{270.6570, 270.0270, 190.4970}

 ] 

\noindent High-gain parametric down conversion is a basic tool applied for preparation of 
quadrature \cite{Wu1986} and two-mode squeezed states \cite{Kumar90}, as well as macroscopic Bell states \cite{Iskhakov2011}. All these examples are considered as macroscopic quantum states of light, which are particularly appealing for providing stronger interaction with matter than microscopic (single-photon and few-photon) states of light. Gravitational wave detection \cite{McKenzie2002}, quantum memory \cite{Polzik2006} and super resolution \cite{Dowling2009} are the other possible applications of macroscopic states of light. However, even though huge progress has been made utilizing high-gain parametric down conversion there is still a number of issues to uncover. The current paper concerns itself with the mode structure of the high-gain parametric down conversion.

For the biphoton state generated via spontaneous parametric down conversion (PDC) it is a common method to apply the Schmidt decomposition of a bipartite system $|\Psi\rangle = \sum_{n=0}^{\infty}\sqrt{\widetilde{\lambda}_{n}}|\psi_{n}\rangle|{\phi_{n}}\rangle$ to a biphoton wavefunction \cite{Law2004} to obtain degree of entanglement represented by the Schmidt number $\widetilde{K} = \frac{1}{\sum_{n=0}^{\infty}\widetilde{\lambda}_{n}^{2}}$. Here the Schmidt coefficients $\widetilde{\lambda}_n$ are independent on the parametric gain. The traditional technique to determine Schmidt number is a Fedorov ratio measurement \cite{Fedorov2004},\cite{Straupe2007}. However it has been shown theoretically \cite{Chan07} that this procedure lacks invariance under translating the detection apparatus from the near field to the far field zone. A direct Schmidt number measurement for the biphoton based on two-photon interference visibility was suggested in same work, and was recently realized experimentally in \cite{Just2013}.

In this letter we experimentally demonstrate an invariant estimation method for the Schmidt number of the high-gain PDC state of light using a more direct detection technique. Efficient single-mode bright squeezed vacuum generation is still a nettlesome problem and we believe that our work will provide a reliable method to quickly characterize a high-gain parametric down conversion state in terms of Schmidt modes.

In contrast to biphoton fields, the representation of bright squeezed vacuum state in terms of wave function is more complicated. Therefore we describe it in the Heisenberg picture using Bloch-Messiah decomposition. In this case one can introduce new photon creation (annihilation) operators in so called broadband modes~\cite{Mau09}. The Schmidt number $K$ is defined by the weights $\lambda_n$ of new the broadband modes and, therefore, depends on the parametric gain~\cite{Per14}

\begin{equation}
K=\dfrac{1}{\sum_{n=0}^{\infty}{\lambda_{n}^{2}}},
\label{K}
\end{equation}

\begin{equation}
\textnormal{and}~\lambda_{n}=\dfrac{\sinh^{2}(\sqrt{\widetilde{\lambda}_{n}} G)}{N}.
\label{lambda}
\end{equation}

\noindent Here $G$ is a parametric gain, $N=\sum_{n=0}^{\infty}\sinh^{2}(\sqrt{\widetilde{\lambda}_{n}} G)$ is the total number of photons in one beam, and $\widetilde{\lambda}_{n}$ are initial  Schmidt coefficients in the decomposition of the biphoton amplitude. The Schmidt number defined in (\ref{K}) gives the effective number of active modes of the the infinite-dimensional bipartite bright squeezed vacuum state and can be obtained experimentally via correlation function measurement~\cite{Ivanova2006}.

The second-order intensity cross-correlation function $g^{(2)}_{si}$ for the signal and idler beams decreases with the number of modes $K$ and the number of photons per mode $N_{mode}$ according to the law

\begin{equation}
g^{(2)}_{si}=1+\frac{1}{K}+\frac{1}{N_{mode}K}.
\label{g2si}
\end{equation}

Considering only signal or idler beams, one can obtain that the correlation function $g^{(2)}$ depends only on $K$ but not depend on the brightness of the state~\cite{Ech11}

\begin{equation}
g^{(2)}=1+\frac{1}{K},
\label{g2ss}
\end{equation}

\noindent i.e., it is sufficient to measure $K$ in one of the beams to describe the whole bipartite PDC state. Robustness against detection losses makes the correlation function measurement a strong tool for the Schmidt number estimation. Surprisingly the applicability of the well known technique (\ref{g2ss}) has never been experimentally validated over the propagation from far to near field regions. In this work we fill up this gap and prove the suitability of the approach at any position of the detection scheme.

\begin{figure} [h]
\centerline{\includegraphics[width=8cm]{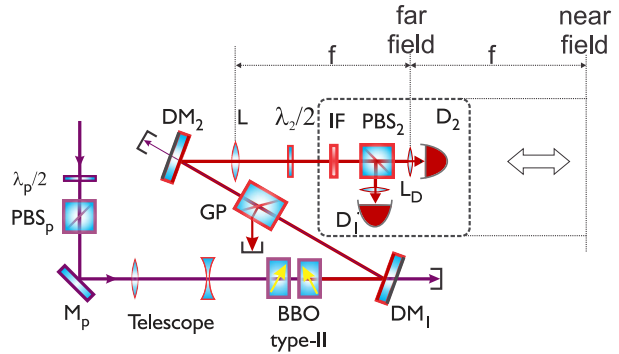}}
\caption{(Color online) The experimental setup.}
\label{fig:1}
\end{figure}

The experimental setup is shown in Figure \ref{fig:1}. Bright Squeezed vacuum beams were produced in an optical parametric amplifier (OPA) by pumping it with the 18-ps pulses of the third harmonic of a Nd:YAG laser at 355 nm wavelength, with a repetition rate 1 kHz. The laser beam diameter was reduced to 230 $\mu$m by a telescope based on a convex lens ($f=50 cm$) and a concave lens ($f=-7.5$) separated by a distance of $42.5 cm$. The OPA consisted of two 5-mm thick BBO crystals cut for collinear degenerate type-II phase matching separated by a distance of 1 cm. Effect of anisotropy was compensated~\cite{Perez2013} by orienting the optical axes (yellow arrows) in the horizontal plane with the directions as shown in the figure. 
PDC radiation was separated from the pump beam by two dichroic mirrors ($DM_{1,2}$) which were transparent for 355 nm and had high reflectivity at 709 nm. After the mirrors the pump beams were absorbed by the beam blocks. Assuming the orthogonal polarization of the PDC beams, the idler one (ordinary polarized) was eliminated by a Glan prism (GP). The frequency spectrum and, therefore, the angular distribution of the PDC radiation was restricted by 10 nm bandwidth interference filter (IF) around 710 nm. The correlation function was measured in a Hanbury Brown-Twiss (HBT) interferometer in which the beam was split in two by means of a half-wave plate ($\lambda_2/2$) and a polarization beamsplitter ($PBS_2$) and then detected by two separate p-i-n diode photodetectors ($D_{1,2}$). A detailed description of the detectors and the registration part of the setup can be found in~\cite{ourPRL2009}.

In our measurement we collected the signals per pulse over 30000 pulses and the correlation function was calculated according to the formula $g^{(2)}=\frac{\langle S_1 S_2 \rangle}{\langle S_1 \rangle\langle S_2 \rangle}$. Here $S_{1,2}$ were detected signals per pulse. The interferometer was built on a translation stage (shown by a dashed rectangle) which was moved from the far-field to near-field zones of a lens (L) with a focal distance of 15 cm. We used a configuration $2f-2f$ where the lens was installed in a double focal distance from the output surface of the second crystal. Therefore the far-field zone was at the focal plane of the lens while the near field was at twice the focal distance. The key issue in the experiment is that the whole angular spectrum was always (at any position of the interferometer) focused onto the detectors by means of plano-convex spherical lenses ($L_D$) with focal length 3 cm.

Based on the measured average signal of the parametric down conversion as a function of the pump power the parametric gain of the optical parametric amplifier was estimated. The pump power was varied in the range from $13$ mW to $25$ mW and the gain took the values within the range $5.8 < G < 8.0$.

\begin{figure} [h]
\centerline{\includegraphics[width=9cm]{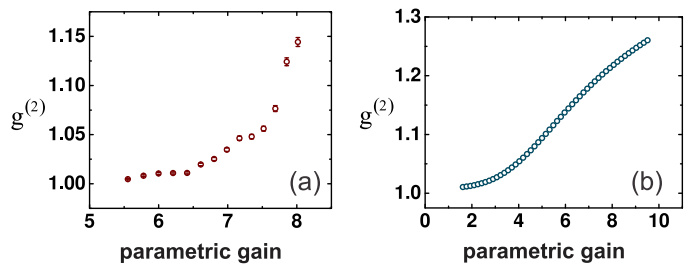}}
\caption{(Color online) $g^{(2)}$ dependence on the parametric gain: (a) Experimentally measured data. (b) Result of the numerical calculation.}
\label{g2vsgain}
\end{figure}

A first step was to measure a correlation function for the extraordinary polarized (signal) PDC beam versus the parametric gain. 
The gain was changed by adjusting pump power with a half-wave plate ($\lambda_p/2$) followed by a polarization cube ($PBS_p$). As shown in Figure~\ref{g2vsgain} (a) we observed the increase of $g^{(2)}$ (the reduction of the number of modes) with $G$. As reported in Fig.~\ref{g2vsgain}~(b), qualitatively similar dependence was obtained numerically. The simulation was performed according to formulas (\ref{K}-\ref{lambda}, \ref{g2ss}) taking into account the parameters of the experiment under the assumption of a single frequency mode. One observes a tendency of the OPA to emit spatially single-mode radiation at high parametric gain. Assuming the obtained results and sensitivity $g^{(2)}$ to the pump power, all further measurements were performed at a constant pump power $P=20.5 mW$, which corresponds to  a parametric gain of $G=7.3\pm0.2$. 


\begin{figure} [h]
\centerline{\includegraphics[width=6cm]{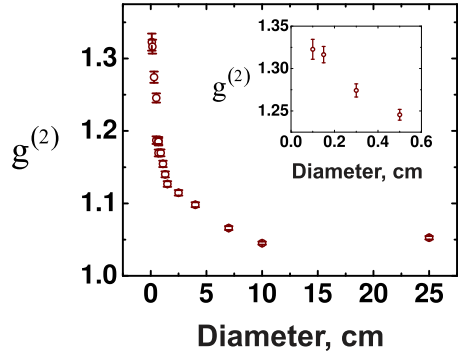}}
\caption{(Color online) $g^{(2)}$ measured as a function of the aperture diameter. Red circles are the experimentally measured points. Inset shows a zoomed part of the dependance obtained with small apertures.}
\label{g2vsAp}
\end{figure}

As a next step, we measured the number of temporal modes in a signal beam containing only one spatial mode. The $g^{(2)}$ dependence on the aperture size placed in the focal plane of the lens (L) is presented in Fig.~\ref{g2vsAp}. The aperture diameter $D$ was changed in the range of $0.1~mm <D<2.5~mm$. It is known that the total Schmidt number $K \equiv K_{t}\cdot K_{s}$ is given by a product of the number of temporal $K_{t}$ modes and spatial $K_{s}$ ones.  We would like to point out, that spatial filtering with a pinhole much smaller than the coherence area results in a lossy  single-mode selection. As one can see in the inset of Fig.~\ref{g2vsAp}, the correlation function tends to $g^{(2)}=1.32\pm0.01$ as $D\rightarrow 0$. According to (\ref{g2ss}) the effective number of temporal modes was found to be $K_{t}=3.1\pm 0.1$.

\begin{figure} [h]
\centerline{\includegraphics[width=9cm]{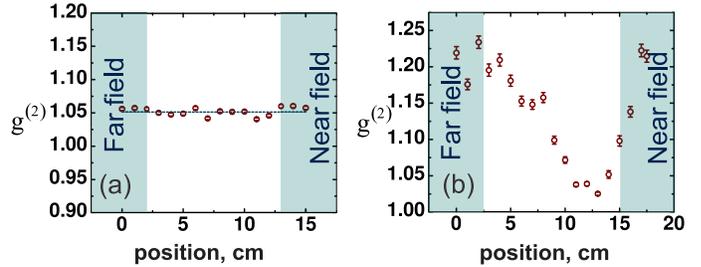}}
\caption{(Color online) Second-order intensity correlation function measured as a function of the position of the detection scheme: (a) All the angular spectrum was collected. (b) An aperture of diameter $D=0.1~cm$ was introduced in front of $PBS_2$. Red circles are the experimentally measured points in agreement
with the theoretical expectation (dashed line).}
\label{g2vsdistance}
\end{figure}

As a confirmation of the applicability of the Schmidt number estimation technique we present $g^{(2)}$ measured at different positions in the region from far to the near field zones. The results are shown by red circles in Fig.~\ref{g2vsdistance}~(a).  In this measurement the pinhole was removed and the whole angular spectrum of the signal beam was collected. One can see that $g^{(2)}$ and, therefore $K$, does not change over the propagation from far field to the near field zone. On average the second-order correlation function took the value of $g^{(2)}=1.052\pm 0.001$. Hence, according to (\ref{g2ss}), we get a total Schmidt number of $K=19.2\pm0.4$. Taking into account $K_t=3.1\pm0.1$ as measured before, we obtain $K_s=6.2\pm 0.2$. For comparison, we show a theoretical expectation by a dashed line. The theoretical value of the spatial Schmidt number $K_s^{(theor)}=6.18$ was calculated for the used OPA operating at a gain of 7.3.

In the last measurement we have installed a 5 mm diameter pinhole in front of $PBS_2$. As shown in Fig.~\ref{g2vsdistance}~(b), in the Fourier conjugated zones the pinhole almost selected a single spatial mode and, therefore, $g^{(2)}$ achieved high values. As expected, in the intermediate zone all the transverse modes become spatially overlapped and all contribute to the signal passing through the pinhole. Hence, the aperture does not select spatial modes any more and the correlation function reaches its minimum. This result is very similar to a dependance of Fedorov ratio on the coordinate recently obtained for biphotons in~\cite{Just2013}.


In conclusion, we have shown, both theoretically and experimentally, that the Schmidt number for the bright squeezed vacuum state decreases with increasing parametric gain. Our results demonstrate that correlation function measurement represents a useful tool for the description of the multi-dimensional quantum systems. The observed invariance of the correlation function on the position of the detection scheme is an experimental verification of the applicability of the Schmidt number estimation technique. We believe that this approach will significantly simply the analysis of bipartite macroscopic entangled states of light.

The would like to acknowledge Stanislav Straupe and Felix Just for fruitful discussions and help in preparing the manuscript. 

\end{document}